\documentclass[fleqn,usenatbib,usedcolumn]{mnras}

\usepackage{graphicx}
\usepackage{amsmath}
\usepackage{amssymb}
\usepackage{bm}
\usepackage{ulem}

\newcommand{\fR}{\bar{f}_\mathrm{R0}}

\title[Rotation Curves in $f(R)$ Gravity]{Imprints of Chameleon $f(R)$ Gravity on Galaxy Rotation Curves}

\author[A. P. Naik et al.]{Aneesh P. Naik,$^{1}$\thanks{E-mail: an485@ast.cam.ac.uk}
Ewald Puchwein,$^{1,2}$
Anne-Christine Davis,$^{3}$ and
Christian Arnold$^{4}$
\\
$^{1}$Institute of Astronomy, University of Cambridge, Madingley Road, Cambridge, CB3 0HA, UK\\
$^{2}$Kavli Institute for Cosmology, University of Cambridge, Madingley Road, Cambridge, CB3 0HA, UK\\
$^{3}$Department of Applied Mathematics and Theoretical Physics, Centre for Mathematical Sciences, Cambridge CB2 0WA, UK\\
$^{4}$Institute for Computational Cosmology, Department of Physics, Durham University, South Road, Durham DH1 3LE, UK
}

\date{Accepted XXX. Received YYY; in original form ZZZ}
\pubyear{2017}

\begin{document}
\label{firstpage}
\pagerange{\pageref{firstpage}--\pageref{lastpage}}
\maketitle


\begin{abstract}

Current constraints on gravity are relatively weak on galactic and intergalactic scales. Screened modified gravity models can exhibit complex behaviour there without violating stringent tests of gravity within our Solar System. They might hence provide viable extensions of the theory of gravity. Here, we use galaxy kinematics to constrain screened modified gravity models. We focus on chameleon $f(R)$ gravity and predict its impact on galaxy rotation curves and radial acceleration relations. This is achieved by post-processing state-of-the-art galaxy formation simulations from the \textsc{auriga project}, using the \textsc{mg-gadget} code. For a given galaxy, the surface dividing screened and un-screened regions adopts an oblate shape, reflecting the disc morphology of the galaxy's mass distribution. At the `screening radius'---the radius at which screening is triggered in the disc plane---characteristic `upturns' are present in both rotation curves and radial acceleration relations. The locations of these features depend on various factors, such as the galaxy mass, the concentration of the density profile and the value of the background field amplitude $\fR$. Self-screening of stars and environmental screening also play a role. For Milky Way-size galaxies, we find that a model with $|\fR|=10^{-7}$ results in rotation curves that are indistinguishable from $\Lambda$CDM, while for $|\fR| \geq 2 \times 10^{-6}$ the simulated galaxies are entirely unscreened, violating Solar System constraints. For intermediate values, distinct upturns are present. With a careful statistical analysis of existing samples of observed rotation curves, including lower mass objects, constraints on $f(R)$ gravity with a sensitivity down to $|\fR|\sim10^{-7}$ should be possible.

\end{abstract}

\begin{keywords}
dark energy -- methods: numerical -- cosmology: theory -- galaxies: general
\end{keywords}

\section{Introduction}
\label{S:Intro}

At the time of writing, nearly two decades have passed since the discovery \citep{Perlmutter1998, Riess1998} that our Universe is undergoing an accelerated expansion. In the intervening period, high precision experiments such as the Planck mission \citep{Planck2016} have found remarkably good agreement with the $\Lambda$CDM model of the Universe. However, underlying uncertainties about the nature of dark matter and and the origin of a cosmological constant provide motivation to think about alternatives and extensions to the theory. It is also not clear whether gravity deviates from pure general relativity theory only at the level of the background expansion of the Universe or whether structure formation on other scales is also altered. As current tests of gravity on cluster, galaxy and intergalactic scales are relatively weak, such modifications could have evaded detection.

Screened modified gravity theories provide a range of viable extensions to GR and $\Lambda$CDM. They exhibit modifications to the gravitational force on different scales \citep[see, e.g.,][]{Joyce2015, Burrage2018, Clifton2012, Koyama2016} while remaining consistent with stringent tests of gravity performed within our Solar System \citep[see, e.g., the review of Solar System constraints in][]{Sakstein2018}.

The subclass of modified gravity (henceforth MG) theories investigated in this work is $f(R)$ gravity. First studied in \citet{Buchdahl1970}, this theory modifies the Einstein-Hilbert action underpinning General Relativity (GR), so that the Ricci scalar $R$ is replaced with a function $R+f(R)$. 

The functional form for $f(R)$ employed in this work is that of \citet{Hu2007}, which is an example of a theory known to exhibit a chameleon screening mechanism \citep{Khoury2004, Brax2008}; the mass of the scalar field is environment-dependent, such that fifth forces are suppressed in regions of high density or in deep potential wells. It is worth also noting that the speed of gravitational waves in $f(R)$ gravity is equal to the speed of light, so the theory remains consistent with the constraints of GW170817 \citep{Brax2016, Lombriser2017, Sakstein2017, Maria2017}. The details of the theory will be outlined in greater detail in \S\ref{S:Theory}.

The key physical parameter to be constrained in this theory is the background amplitude of the scalar field at redshift zero, $|\fR|$. This controls where screening is triggered and hence the magnitude of modified gravity effects in different environments. Smaller $|\fR|$ values corresponding to smaller deviations from GR. An overview of constraints on chameleon theories, including chameleon $f(R)$ theories, can be found in \citet{Burrage2018}. Currently, the strongest constraints on $|\fR|$ come from tests using cosmic distance indicators. The distances inferred from observations of Cepheid variables and stars at the tip of the red-giant branch would be affected if these objects are exposed to a modified gravitational force. \citet{Jain2013} derive a 95\% confidence limit of $|\fR| < 5 \times 10^{-7}$ from such observations. Other competitive constraints, at around the $10^{-6}$ level, have been calculated from redshift-space distortions \citep{Xu2015}, and by comparing the stellar and gaseous rotation curves of dwarf galaxies \citep{Vikram2014}. Similarly, strong constraints can be obtained by requiring that modified gravity effects are screened in the Milky Way at the solar radius, although this depends on the amount of environmental screening of the Milky Way by the Local Group. To make further progress on constraining modified gravity on the scale of galaxies, more accurate predictions of how modifications of gravity affect the kinematics of galaxies are needed.  

Hitherto, most works investigating the effect of modifications of gravity on galactic scales have done so with the aim of replacing the role of dark matter, with theories such as MOND \citep{Milgrom1983}. One recent work was \citet{Burrage2017}, in which it was found that a symmetron fifth force can provide a good fit to observed radial acceleration relations, in place of dark matter. The present work, in contrast, studies the kinematics of galaxies in MG theories that assume a similar amount of dark matter as in $\Lambda$CDM but alter structure formation by a fifth force. Little work has been done in this direction previously. Notable exceptions include \citet{Almeida2018} and \citet{Vikram2014}. The former introduces a Yukawa-like fifth force with a different coupling to dark matter and baryons. The study of \citet{Vikram2014} will be discussed in more detail in \S\ref{S:Discussion}.

Another notable study constraining modified gravity with galaxies is \citet{Desmond2018b}. They constrain fifth forces, including those of chameleon $f(R)$ theories, in the local cosmic web by searching for offsets between the gas and stellar centroids of galaxies, deriving a tight constraint of $|\fR| \leq \mathrm{few} \times 10^{-8}$.

The present work, by contrast with all of the above, investigates the effect of screened chameleon $f(R)$ gravity on galaxy rotation curves, providing the first simulation predictions of rotation curves in screened modified gravity.

N-body simulations of alternative gravity theories are increasingly the tool of choice for predicting their observational signatures. \citet{Winther2015} present an overview and comparison of different simulation codes and find good agreement between the results of different groups, indicating that the field has reached a significant degree of maturity. \citet{Arnold2016} recently performed the first high resolution, dark matter-only, zoom-in simulations of Milky Way-sized halos in $f(R)$ gravity. This work builds upon that work by providing the first simulated galaxy rotation curves in $f(R)$ gravity for Milky Way-sized galaxies.  

A key difference worth emphasising between the work of \citet{Arnold2016} and the present work is that the former work provides fully self-consistent solutions of $f(R)$ dark matter halos. The halos simulated therein are evolved under $f(R)$ gravity from high redshift initial conditions in the linear regime to the present day. The extreme non-linearity of the governing equations of $f(R)$ gravity make these simulations significantly more computationally expensive than in standard gravity. Nevertheless, dark matter-only simulations such as those found in \citet{Arnold2016} are computationally feasible. Simulating the complex hydrodynamics and baryonic physics of galaxies at high resolutions and with modified gravity is, however, still very challenging at present. So, in order to investigate the effects of chameleon $f(R)$ gravity in galaxies, we have instead performed calculations of the $f(R)$ effects in `post-processing'. 

We have used the mass distributions of state-of-the-art $\Lambda$CDM simulations of disc galaxy formation from the \textsc{auriga project} \citep{Grand2017}. Then, using the modified gravity solver aboard the \textsc{mg-gadget} code, the scalar fields and fifth forces were numerically computed across these galaxies, and the corresponding rotation curves derived. As a result, our models are not full dynamical models encapsulating the evolution of $f(R)$ galaxies, but simply rotation curves calculated for galaxies with mass distributions believed to be closely resembling those of real galaxies. The basic underlying assumption is that disc galaxies should have a rotationally-supported disc component even in the presence of modifications of gravity, so that the radial gravitational acceleration in the disc plane can be inferred from measurements of gas and/or stellar velocities, which allows the presence of a fifth force to be constrained.
	
This work is structured as follows. Firstly, \S\ref{S:Theory} outlines some of the theoretical foundations of chameleon $f(R)$ gravity necessary for the subsequent discussion. \S\ref{S:Methodology} then describes the methodology of the present work, describing both the code employed and the simulated galaxies analysed herein. Results are then presented in \S\ref{S:Results} and discussed in \S\ref{S:Discussion}, followed by concluding remarks in \S\ref{S:Conclusions}. The metric signature adopted in this work is $(-, +, +, +)$, and units are chosen such that the speed of light $c=1$.

\section{Chameleon $f(R)$ Gravity}
\label{S:Theory}

The species of MG employed in this work is `$f(R)$ gravity'. $f(R)$ theories were first studied in \citet{Buchdahl1970}, and a review of more recent progress can be found in \citet{Amendola2010}. $f(R)$ theories have been shown to be equivalent to scalar-tensor theories of gravity \citep{Brax2008}, so that constraints on $f(R)$ gravity can be translated into the language of scalar tensor gravity.

The action of this theory is given by:
\begin{equation}\label{E:f(R)Action}
S = \int d^4x\sqrt{-g}\frac{1}{16\pi G}\left[R+f\left(R\right)\right] + S_{m}[g_{\mu\nu},\psi_{i}]
\end{equation}
As the name of the theory would suggest, the Ricci scalar $R$ in the classical Einstein-Hilbert action is replaced with a generalised $R+f\left(R\right)$. The theory reduces to $\Lambda$CDM in the case that $f(R)=-2\Lambda$.

It is worth noting that there is some variation in the literature regarding the definition of $f(R)$. In some places, the convention adopted is $R \rightarrow R+f\left(R\right)$, while elsewhere $R \rightarrow f\left(R\right)$ is used. In this work, the former convention has been adopted in order to maintain consistency with \citet{Puchwein2013} and the internal workings of \textsc{mg-gadget}. 

In order to ensure that a theory of modified gravity passes Solar System constraints, it is necessary to introduce a `screening mechanism', i.e. a mechanism that suppresses the fifth force in dense environments like those of our Solar System, but (for interesting, testable theories) unleashing a fifth force in other environments. A review of screening mechanisms can be found in \citet{Joyce2015}.

Extremising the action in Eq.~(\ref{E:f(R)Action}) leads to an equation of motion for $f_\mathrm{R} \equiv \frac{\mathrm{d}f}{\mathrm{dR}}$, which plays the role of a scalar field in this theory. Furthermore, the acceleration of a free-falling particle in terms of the Newtonian potential $\phi$ and $f_\mathrm{R}$ is given by 
\begin{equation}\label{E:ScalarFieldForceLaw}
\bm{\ddot x} + \bm{\nabla}\Phi = \frac{1}{2}\bm{\nabla}f_\mathrm{R}.
\end{equation}
The fifth force contribution on the right hand side of this equation is directly related to the gradient of the scalar field $f_\mathrm{R}$. The theory can exhibit the screening mechanism known as `chameleon' screening if $f(R)$ is suitably chosen such that $f_\mathrm{R}$ gradients are strongly suppressed in regions of high density, rendering the fifth force undetectable.

A widely studied $f(R)$ model known to exhibit chameleon screening is the Hu-Sawicki model \citep{Hu2007}
\begin{equation}\label{E:HuSawicki}
f(R) = -m^2\frac{c_1\left(\frac{R}{m^2}\right)^n}{1+c_2\left(\frac{R}{m^2}\right)^n},
\end{equation}
where $m^2\equiv H_0^2\Omega_m$; $H_0$ is the present-day value of the Hubble parameter and $\Omega_m$ is the current matter density fraction in units of the critical density. Moreover, \textsc{mg-gadget} follows much of the literature about the Hu-Sawicki model in adopting $n=1$.

The model can recover an expansion history close to $\Lambda$CDM if it is required that $\frac{c_2 R}{m^2} \gg 1$, so that
\begin{equation}
f(R) \approx -m^2\frac{c_1}{c_2}\left[1 + \mathcal{O}\left( \frac{m^2}{c_2 R}
\right) \right].
\end{equation}
Then, one recovers $\Lambda$CDM to first order, i.e. $f(R) \approx -2\Lambda$, if
\begin{equation}
\frac{c_1}{c_2} = 6\frac{\Omega_\Lambda}{\Omega_m}.
\end{equation}
With this, we are left with a free choice of just one parameter: either $c_1$ or $c_2$. 

Differentiating the Hu-Sawicki model in Eq.~(\ref{E:HuSawicki}), the scalar field $f_\mathrm{R}$ is given by
\begin{equation}\label{E:ScalarFieldParameters}
f_\mathrm{R} \equiv \frac{\mathrm{d}f}{\mathrm{d}R} = -\frac{c_1}{\left(\frac{c_2R}{m^2}+1\right)^2} \approx - \frac{c_1}{c_2^2}\left(\frac{m^2}{R}\right)^2
\end{equation}

Using the Friedmann-Robertson-Walker metric, an expression for the background curvature as a function of scale factor $a$ can be derived,
\begin{equation}\label{E:BackgroundCurvature}
\bar{R}(a) = 3\frac{m^2}{a^3}\left(1 + 4\frac{\Omega_\Lambda a^3}{\Omega_m}\right).
\end{equation}
Combining Eqs.~(\ref{E:BackgroundCurvature}) and (\ref{E:ScalarFieldParameters}) yields the following expression for $\fR$, the background value of the present-day scalar field,
\begin{equation}\label{E:BackgroundFieldParameters}
\fR = -\frac{2\Omega_\Lambda \Omega_m}{3\left(\Omega_m + 4\Omega_\Lambda\right)^2}\frac{1}{c_2}.
\end{equation}
This shows that there is only one free parameter remaining in the model. From Equation~(\ref{E:BackgroundFieldParameters}), it is apparent that choosing $c_2$ is equivalent to choosing $\fR$. The $\Lambda$CDM+GR limit corresponds to the limit $\fR\rightarrow 0$ or $c_2 \rightarrow \infty$. In the remainder of this work we purely use $\fR$ as the parameter defining our Hu-Sawicki models.

A final note in this section is that combining Eqs.~(\ref{E:BackgroundCurvature}) and (\ref{E:ScalarFieldParameters}) also leads to expressions for the time-dependent background scalar field $\bar{f}_\mathrm{R}(a)$ and the curvature perturbation $\delta R \equiv R - \bar{R}(a)$ as a function of scale factor $a$,
\begin{equation}\label{E:BackgroundFieldTimeDependence}
\bar{f}_\mathrm{R}(a) = a^6\fR\left(\frac{1 + 4\frac{\Omega_\Lambda}{\Omega_m}}{1 + 4\frac{\Omega_\Lambda a^3}{\Omega_m}}\right)^2,
\end{equation}
\begin{equation}\label{E:CurvaturePerturbation}
\delta R = \bar{R}(a)\left(\sqrt{\frac{\bar{f}_\mathrm{R}(a)}{f_\mathrm{R}}}-1\right).
\end{equation}

Extremising the action in Eq.~(\ref{E:f(R)Action}) gives a set of modified Einstein field equations, which in the Newtonian limit leads to an equation of motion for $f_\mathrm{R}$,

\begin{equation}\label{E:FieldEOM}
\nabla^2 f_\mathrm{R} = \frac{1}{3} \left( \delta R - 8\pi G \delta \rho \right),
\end{equation}
in addition to a modified Poisson equation for the gravitational potential \citep{Hu2007},
\begin{equation}\label{E:ModifiedPoisson}
\nabla^2 \Phi = \frac{16\pi G}{3}\delta\rho -\frac{1}{6}\delta R.
\end{equation}
Here $\delta\rho$ is the perturbation of the matter density from its background value, while $\delta R$ denotes the perturbation of the scalar curvature. Implicit in the derivations of these equations is the assumption $|f_\mathrm{R}| \ll 1$, which is satisfied for all viable models, and the quasistatic approximation $|\nabla f_\mathrm{R}| \gg \frac{\partial f_\mathrm{R}}{\partial t}$. The latter approximation should be well justified in the models considered here, as discussed in \citet{Noller2014} and \citet{Sawicki2015}.

It will prove useful later to rewrite the modified Poisson equation (\ref{E:ModifiedPoisson}) in terms of an `effective density' $\delta\rho_\mathrm{eff} \equiv \frac{1}{3}\delta\rho - \frac{1}{24\pi G} \delta R$ that encodes the modified gravity contributions,

\begin{equation}\label{E:EffectivePoisson}
\nabla^2 \Phi = 4\pi G \left(\delta\rho + \delta\rho_\mathrm{eff}\right).
\end{equation}

Equations (\ref{E:FieldEOM}) and (\ref{E:EffectivePoisson}), as well as (\ref{E:BackgroundFieldTimeDependence}) and (\ref{E:CurvaturePerturbation}) are the governing equations solved by the scalar field solver in \textsc{mg-gadget}.

\section{Methodology}
\label{S:Methodology}

\subsection{The Auriga galaxy formation simulations}
\label{S:Methodology:Auriga}

\begin{table*}
\centering
\begin{tabular}{c | c c c c c c}
Galaxy & $r_{200}$ & $M_{200}$                     & $M_{*}$                       & $v_{200}$ & $v_{\mathrm{peak}}$ \\
       & [kpc]       & [$10^{10}\ \mathrm{M_{\odot}}$] & [$10^{10}\ \mathrm{M_{\odot}}$] & [km/s]      & [km/s]                \\
\hline
Au1	   & 192.09    & 77.38	                       & 3.61                          & 159.44    & 197.19              \\
Au2	   & 260.51    & 193.01                        & 11.19                         & 184.02    & 246.64              \\
Au9    & 232.25    & 104.69                        & 6.26                          & 149.87    & 261.36              \\
Au11   & 212.45    & 149.28                        & 8.67                          & 179.93    & 241.95              \\
Au13   & 239.14    & 122.03                        & 7.30                          & 158.59    & 282.63              \\
Au20   & 223.59    & 127.58                        & 5.89                          & 162.43    & 219.51              \\
Au21   & 226.93    & 146.53                        & 9.13                          & 169.01    & 252.30              \\
Au22   & 237.64    & 91.74                         & 6.20                          & 144.70    & 284.38              \\
Au24   & 203.31    & 147.29                        & 8.11                          & 177.37    & 239.34              \\
AuL1   & 238.07    & 51.70                         & 2.47                          & 120.19    & 165.94              \\
AuL4   & 168.67    & 52.39                         & 2.52                          & 111.91    & 143.13              \\
AuL5   & 184.34    & 68.38                         & 3.43                          & 128.57    & 189.64              \\
AuL8   & 197.83    & 84.52                         & 5.36                          & 140.34    & 223.09
\end{tabular}
\caption[Table of info]{Basic properties of the simulated galaxies used in this work. Columns are 1) Auriga ID: the identifying number of the galaxy in the \textsc{auriga project}, 2) $r_{200}$: the virial radius, here taken as the radius enclosing a region of average density equal to 200 times the cosmic critical density, 3) $M_{200}$: the mass contained within $r_{200}$, 4) $M_{*}$: the total stellar mass within $r_{200}$, 5) $v_{200}$: the circular velocity in the disc plane at $r_{200}$, and 6) $v_{\mathrm{peak}}$: the peak circular velocity.} 
\label{T:Galaxies}
\end{table*}

\begin{figure*}
    \includegraphics[width=1.75\columnwidth]{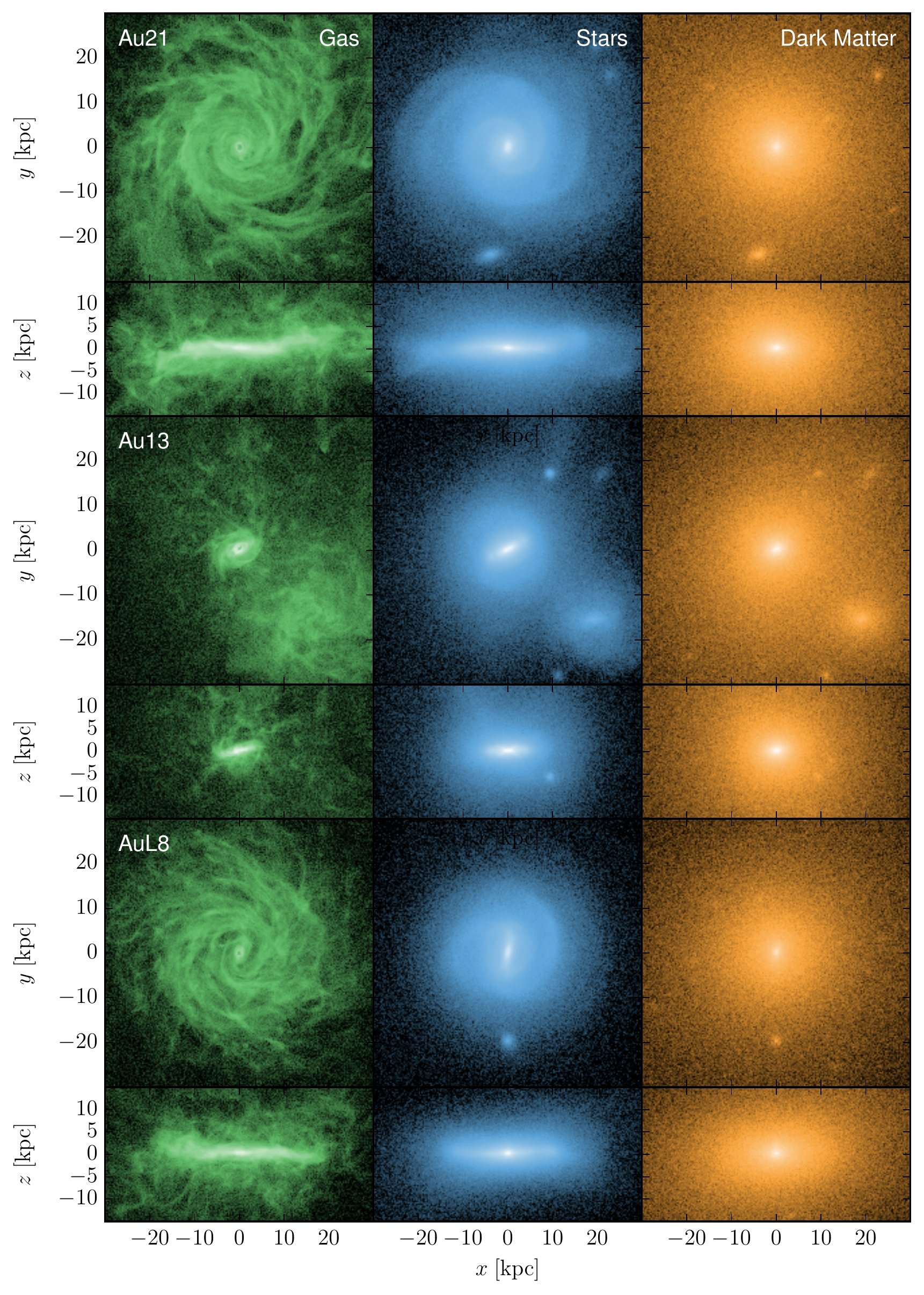}
    \caption{Projected gas (left), stellar (middle) and dark matter (right column) density of the \textsc{auriga} galaxies Au21 (top), Au13 (middle), and AuL8 (bottom row). For each object and component, face-on and edge-on projections are shown. The mass distributions of these and other simulated galaxies were used as an input for the \textsc{mg-gadget} modified gravity solver.}
    \label{F:Galaxies}
\end{figure*}

\begin{figure*}
    \includegraphics[width=2\columnwidth]{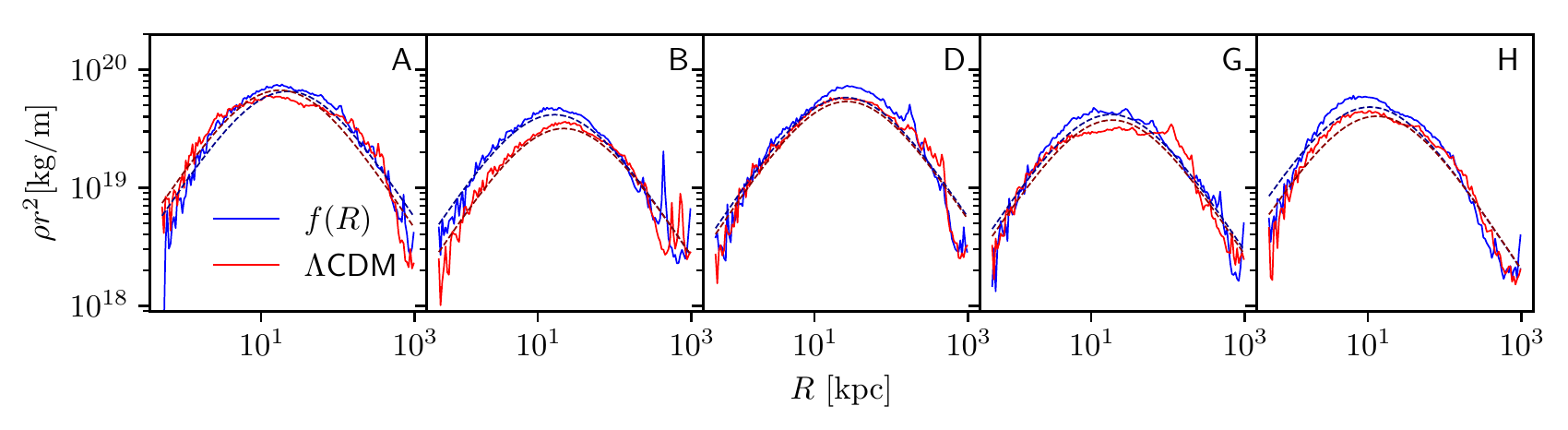}
    \caption{Density profiles of five dark matter haloes. The blue curves represent haloes from the full $f(R)$ simulations, while the red curves are their $\Lambda$CDM counterparts from the original \textsc{aquarius} simulations. The corresponding dashed lines are NFW fits to the density profiles. Density is multiplied by a factor of $r^2$ for improved readability.}
    \label{F:Aquarius_Profiles}
\end{figure*}

\begin{figure*}
    \includegraphics[width=2\columnwidth]{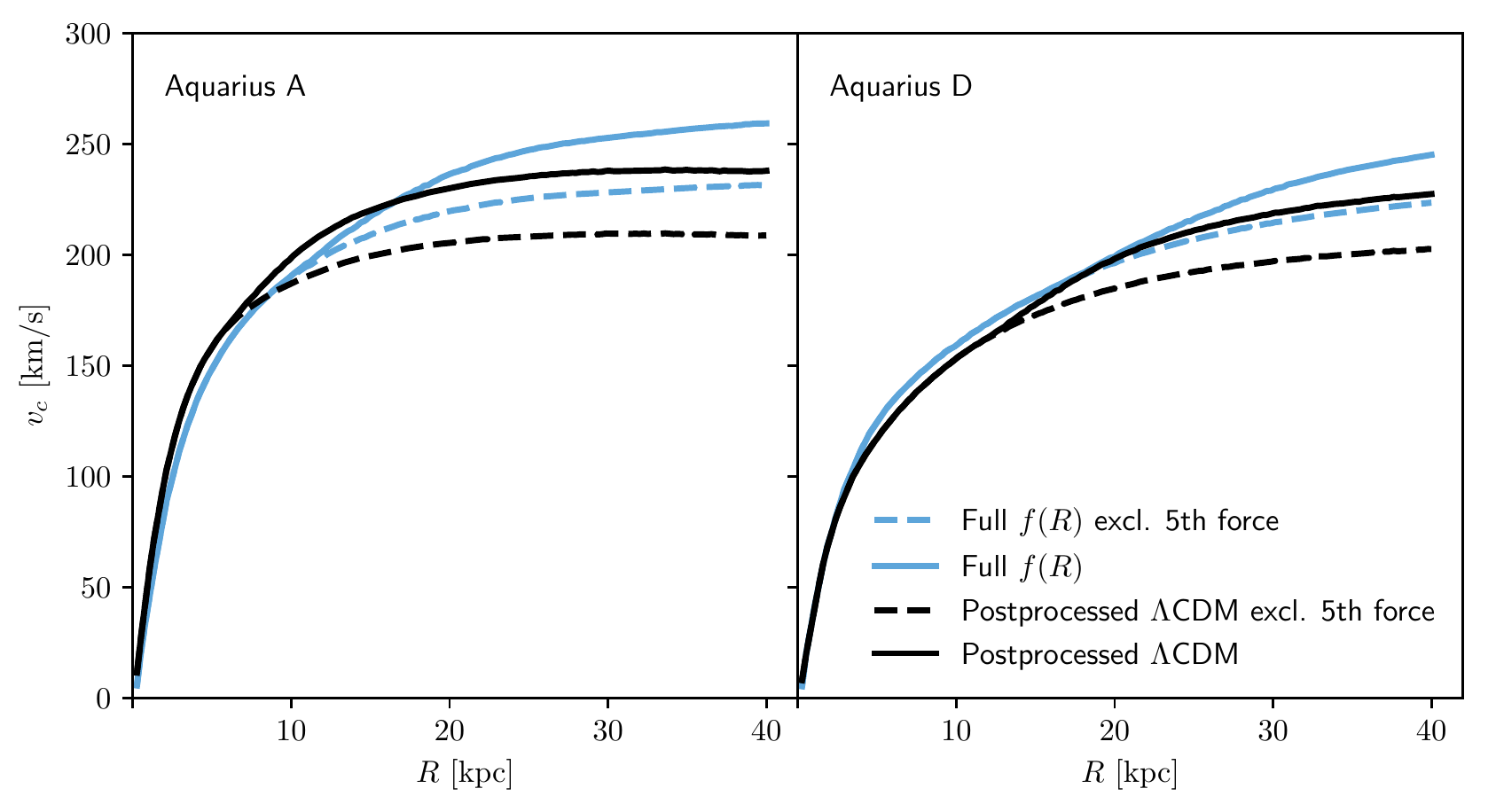}
    \caption{Circular velocity profiles for two \textsc{aquarius} dark matter haloes, labelled A (\textit{left}) and D (\textit{right}). The blue curves show the circular velocity calculated from particle accelerations in the full $f(R)$ simulations, while the black curves are calculated analogously from their post-processed $\Lambda$CDM counterparts from the original \textsc{aquarius} simulations. In all cases, $|\fR| = 10^{-6}$. Solid lines include the fifth force contribution, while the dashed lines ignore it.}
    \label{F:Aquarius_RCs}
\end{figure*}

The simulated galaxies studied in this work were not evolved ab initio using \textsc{mg-gadget}, but were instead formed in hydrodynamical $\Lambda$CDM simulations with state-of-the-art baryonic physics performed in the \textsc{auriga project} \citep{Grand2017}, and post-processed using the scalar field solver of \textsc{mg-gadget}. The validity of this post-processing approach is discussed in \S\ref{S:Methodology:MG-GADGET}.

The \textsc{auriga project} employed magnetohydrodynamics and a sophisticated galaxy formation prescription (including implementations of radiative cooling, star formation, chemical enrichment, supernovae and AGN feedback) to perform zoom simulations of 30 isolated Milky-Way size galaxies, using the moving mesh code \textsc{arepo} \citep{Springel2010}.

The \textsc{auriga} galaxies reproduce a range of observables of Milky Way-like galaxies, including masses, sizes, rotation curves, star formation rates, and metallicities. Furthermore, the simulated galaxies have clear Milky Way-like spiral morphologies; featuring bars and spiral arms. 

Thirteen such galaxies have been studied here, and an overview of their basic properties is provided in Table \ref{T:Galaxies}. 9 of these galaxies are from the original \textsc{auriga} Project (Au1, Au2, Au9, Au11, Au13, Au20 Au21, Au22, Au24), a further 4 lower-mass galaxies (AuL1, AuL4, AuL5, and AuL8) were taken from a follow-up project. 

Projections of the various components, i.e. gas, stellar, and dark matter surface density, of galaxies Au21, Au13, and AuL8 are shown in Figure \ref{F:Galaxies}. These 3 galaxies have been chosen to represent a range of galaxy morphologies. Au21 is a grand design spiral galaxy, AuL8 has a prominent bar, and Au13 a slightly less well-defined disc.

The particle, and hence mass, distributions of these galaxies were extracted from the \textsc{auriga} simulation snapshots, and fed to \textsc{mg-gadget}'s modified gravity solver, which is discussed in the following sub-section and computes the scalar field $f_\mathrm{R}$ and the modified gravity accelerations throughout the simulation volume.

\subsection{Calculation of modified gravity effects}
\label{S:Methodology:MG-GADGET}

\begin{figure*}
    \includegraphics[width=2\columnwidth]{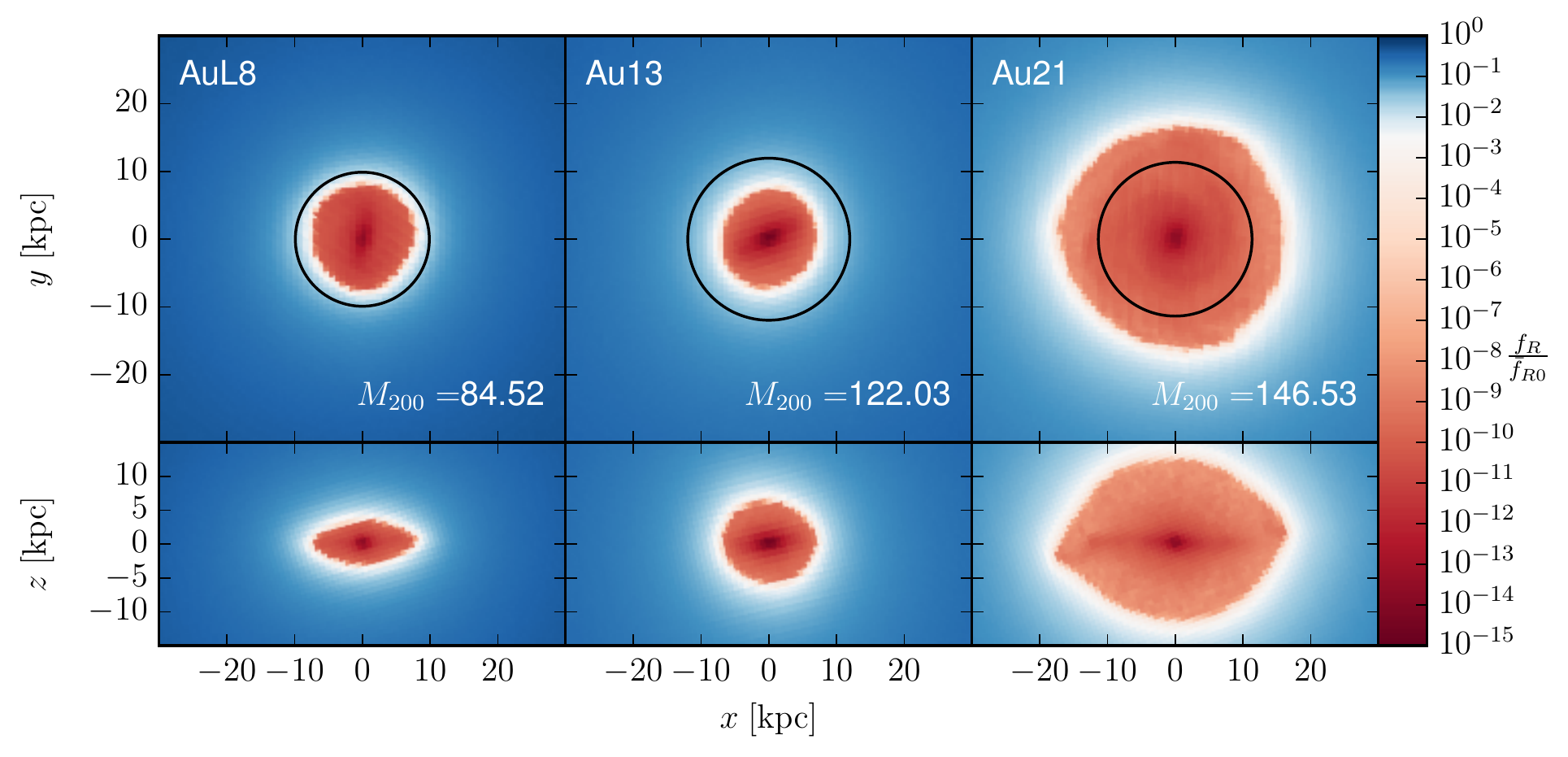}
    \caption{Face-on (\textit{top}) and edge-on (\textit{bottom}) scalar field maps of galaxies AuL8, Au13, and Au21, for a background field amplitude of $|\fR|=8 \times 10^{-7}$. The black circle marks $0.05\, R_{200}$. At the screening surfaces (white regions), the field amplitude drops by many orders of magnitude. The edge-on views demonstrate that the screening surface is typically compressed towards the disc plane, reflecting the density distribution of the galaxy. The virial mass ($M_{200}$) of each galaxy is labelled in each case, in units of $10^{10}M_{\odot}$.}
    \label{F:FieldMaps}
\end{figure*}

\begin{figure*}
    \includegraphics[width=2\columnwidth]{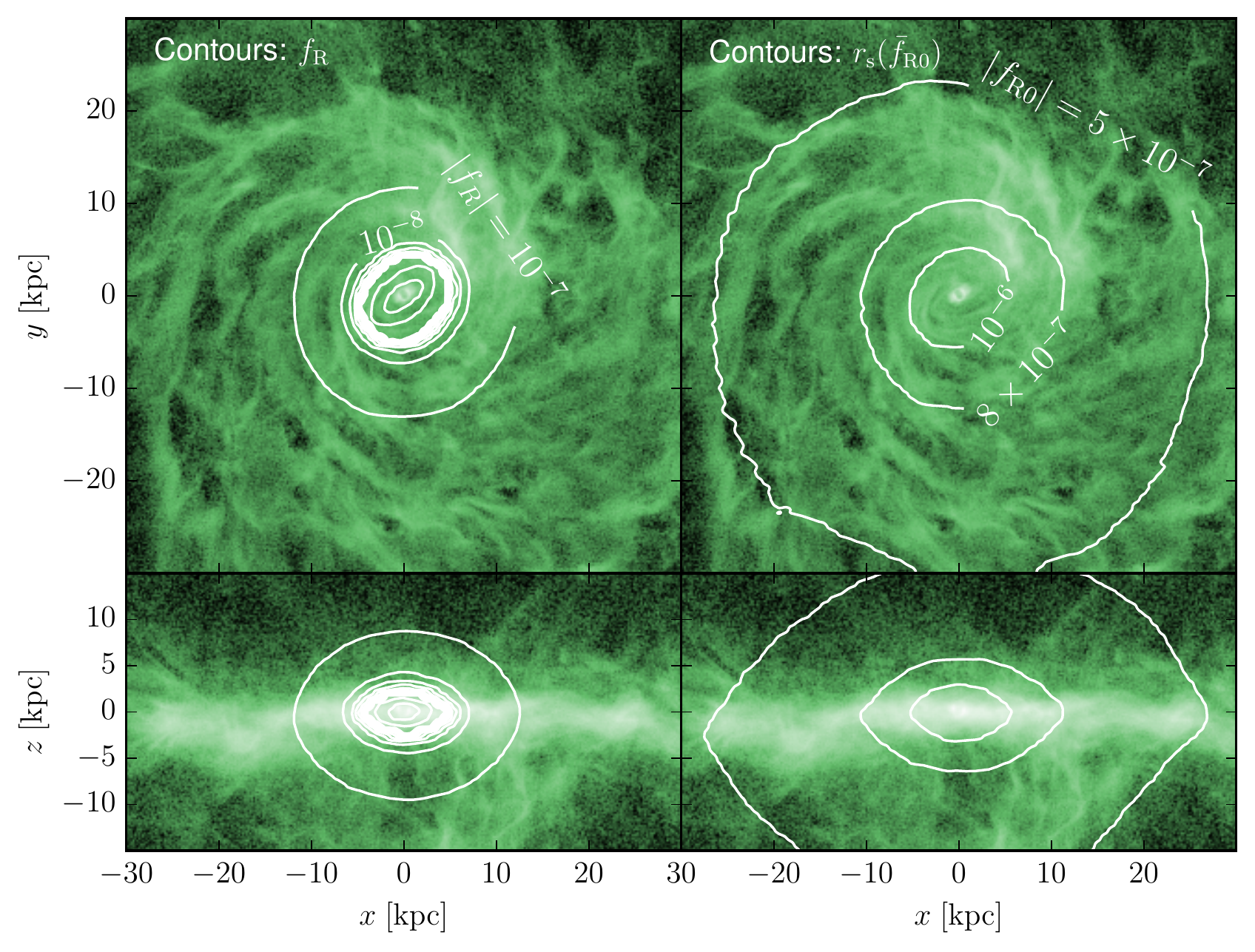}
    \caption{\textit{Left}: Face-on (\textit{top}) and edge-on (\textit{bottom}) contour maps of the scalar field, for Au20 and a background field amplitude of $|\fR|=10^{-6}$, evaluated in planes passing through the galaxy centre. \textit{Right}: Face-on and edge-on contours showing the location of the screening surface for three \textit{different} values of the background field amplitude, $|\fR|$: $10^{-6}$, $8\times 10^{-7}$, and $5\times 10^{-7}$. The screening surface is here defined as the surface on which $f_\mathrm{R} = 10^{-4}\fR$. The contours are superimposed on maps of the projected gas density.}
    \label{F:Contours}
\end{figure*}

The modified gravity solver used in this work is part of the \textsc{mg-gadget} simulation code \citep{Puchwein2013}, which is itself based on the \textsc{p-gadget3} code \citep[last described in][]{Springel2005}, but incorporates a reworked gravity solver. The latter allows simulating models with highly non-linear force laws such as Hu-Sawicki $f(R)$ gravity. 

In the base \textsc{p-gadget3} code, gravitational forces are calculated using a `TreePM' method: long-range forces are calculated using Fourier (Particle Mesh) methods, while short-range forces are calculated using a hierarchical oct-tree, which gives higher spatial resolution. \textsc{mg-gadget} also utilises these methods to solve Eq. (\ref{E:EffectivePoisson}), but in addition the scalar field $f_\mathrm{R}$ is computed and stored on a space-filling adaptive mesh, which is constructed from the oct-tree structure. More precisely, $f_\mathrm{R}$ is obtained by solving Eq. (\ref{E:FieldEOM}) with a multi-grid accelerated, iterative Newton-Gauss-Seidel relaxation method on the adaptive mesh. This allows calculating $f_\mathrm{R}$ everywhere in the simulation volume, as well as the MG acceleration on each particle. A much more detailed description of the algorithm can be found in the original code paper \citep{Puchwein2013}, while scientific applications of the code are presented in, e.g., \citet{Arnold2018} and \citet{Arnold2016}.

The modified gravitational forces are calculated in post-processing from $\Lambda$CDM simulations, rather than by performing full galaxy formation $f(R)$ simulations, which would be computationally much more expensive. This means that modified gravity effects on the evolution of galaxies will not be captured. However, the gas and stellar components of the \textsc{auriga} galaxies match observations well and can thus be used as a mass model for investigating the modified gravitational forces in galaxies in $f(R)$ gravity. Assuming that $f(R)$-galaxies also have a rotationally-supported disc component, the corresponding effects on the rotational velocity can be estimated. An additional uncertainty is introduced by the change in the dark matter density profile due to modified gravity.

\citet{Arnold2016} investigated the effect of chameleon $f(R)$ gravity on the formation of dark matter halos. In that work, computationally cheaper dark matter-only $f(R)$ gravity simulations were performed using \textsc{mg-gadget} in order to investigate the effects of $f(R)$ gravity on Milky Way-mass dark matter halos. The simulations were performed from identical initial conditions to the \textsc{aquarius} simulations \citep{Springel2008}. Figure \ref{F:Aquarius_Profiles} compares the density profiles of five haloes in these full $f(R)$ simulations (for $|\fR|=10^{-6}$) with those of the corresponding haloes in the $\Lambda$CDM simulations. It can be seen in this figure that the shapes of the density profiles from the $f(R)$ simulations are qualitatively similar to those of the $\Lambda$CDM simulations. While the presence of the chameleon fifth force might change the total mass and concentration of a given halo somewhat, it does not significantly affect its morphology. In particular, the NFW profile \citep{Navarro1996} can fit both $\Lambda$CDM and $f(R)$ haloes equally well. When fitting observed rotation curves along with measured stellar and gas densities, the mass and concentration of the dark matter halo would be free parameters, so that changes in these parameters due to modified gravity would be captured.

Figure \ref{F:Aquarius_RCs} shows circular velocity profiles (rotation curves) for two of the five haloes. These two were chosen by virtue of being the only ones partially screened, rather than fully unscreened, for $|\fR|=10^{-6}$. The blue curves show rotation curves from the original full $f(R)$ simulations, while the black curves show post-processed rotation curves from the $\Lambda$CDM simulations. The key point of this figure is that both sets of rotation curves are qualitatively the same, with similar upturns at the screening radii. The differences in the heights of the curves and locations of the screening radii can be ascribed to the differences in halo mass and concentration. That is to say, employing the fully self-consistent approach of full $f(R)$ simulations would have changed the exact masses and density profiles of the galaxies, and therefore the resulting rotation curves, but the qualitative features, particularly the upturns, would still be present at the screening radii.

Another complication is the potential effect of baryonic feedback on the dark matter density profile \citep[see, e.g.,][]{Duffy2010}. In the following we assume that such effects (e.g., the potential formation of a core) happen in a very similar way under both $f(R)$ gravity and $\Lambda$CDM. This should certainly hold in objects in which the central region, which is most prone to baryonic effects, is screened. Using the assumption of similar baryonic effects, it is then possible to describe both $f(R)$ and $\Lambda$CDM halo profiles with the same functional form.

In principle it would be interesting to test the effect of modified gravity on the baryonic feedback. This is, however, very difficult due to the wide range of scales involved and due to our limited understanding of the relevant astrophysics. As an example of how these processes might differ under modified gravity, \cite{Davis2012} find that unscreened stars are typically brighter in modified gravity scenarios, which implies higher supernova rates than in $\Lambda$CDM.

It is also worth noting that baryonic feedback is typically implemented in galaxy formation simulations not via detailed models of all the relevant physics, but via strongly simplified subgrid models which need to be calibrated to observational constraints. Hence, even when evolving such simulations fully under modified gravity, changes in the baryonic effects due to the fifth force would likely be partially offset by re-calibrating the model parameters to observations. Furthermore, as shown in \cite{Davis2012}, this differences between modified gravity and $\Lambda$CDM feedback mechanisms can be expected to be negligible at the small field amplitudes considered in this work $|\fR| \leq 10^{-6}$, at which many galactic baryons inhabit screened regions. Thus, our assumptions should remain robust.

These considerations, together with the assumption that rotationally-supported disc components are present in both chameleon $f(R)$ gravity and in $\Lambda$CDM, suggest that our post-processing approach is valid for illuminating the impact of fifth forces on galaxy rotation curves. That is, qualitatively similar rotation curve upturns and radial acceleration relation bumps would have been seen in the \textsc{auriga} galaxies had they instead been simulated fully self-consistently under $f(R)$ gravity. The exact location of the features might be slightly different due to modified gravity effects on the dark matter density profiles. However, the parameters describing the dark matter profiles would be free parameters when fitting observed rotation curves, so that these effect would be taken into account.

\subsection{Rotation Curves}
\label{S:Methodology:RCs}

\begin{figure*}
    \includegraphics[width=2\columnwidth]{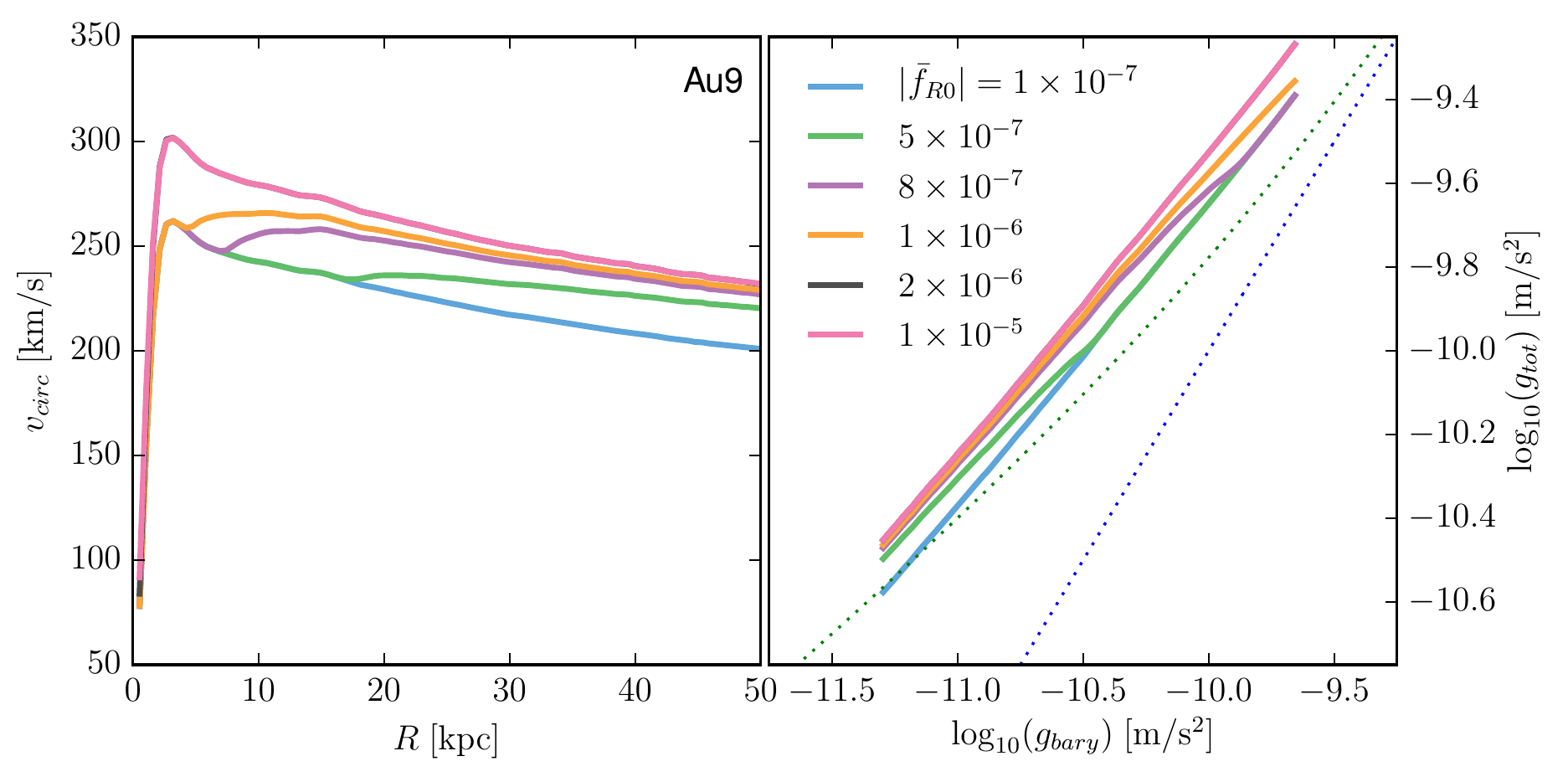}
    \caption{\textit{Left:} Rotation curves of Au9 for six different values of $|\fR|$, as labelled. As described in the text, $v_{\mathrm{circ}}$ is calculated from the full gravitational acceleration, including a potential fifth force. \textit{Right:} Radial acceleration relations for Au9, for the same $|\fR|$ values. The blue dotted line represents $g_{\rm tot}=g_{\rm bary}$, while the green dotted line represents the best-fit function for observed radial acceleration relations from the SPARC sample. $g_{\mathrm{tot}}$ is
    based on the full gravitational acceleration, including a potential fifth force, while $g_{\mathrm{bary}}$ is calculated at each radius from the enclosed baryonic mass assuming spherical symmetry and standard gravity, i.e. using $v_\mathrm{circ} = GM_\mathrm{bary}(<R)/R^2$. Note that in both panels, the line corresponding to $|\fR|=2 \times 10^{-6}$ is obscured behind that of $|\fR|=1 \times 10^{-5}$, as both correspond to fully unscreened cases.}
    \label{F:Au9_RC_RAR}
\end{figure*}

\begin{figure*}
    \includegraphics[width=1.75\columnwidth]{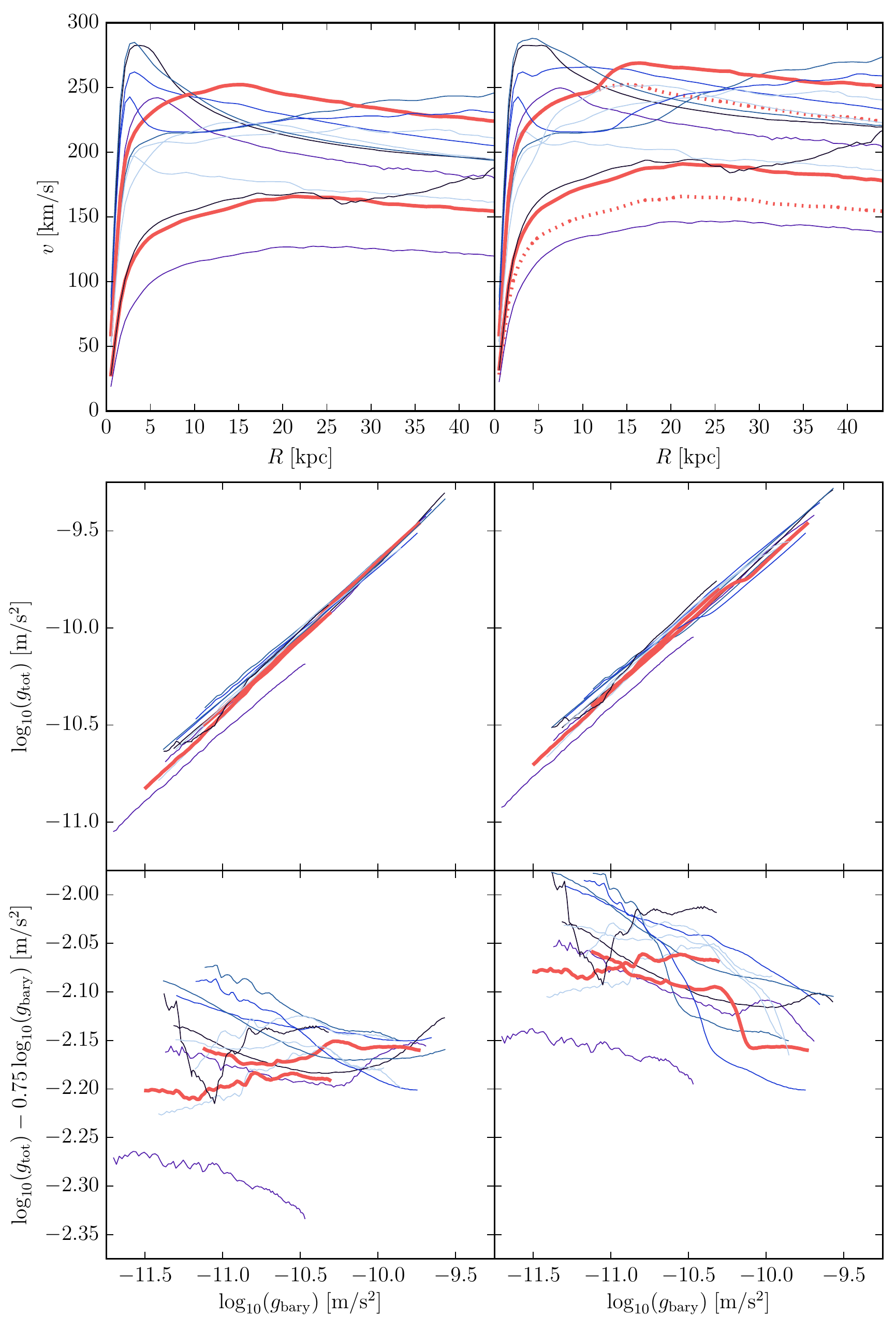}
    \caption{Rotation curves (\textit{top}) and radial acceleration relations (\textit{middle and bottom}) for all 13 galaxies in the $\Lambda$CDM (\textit{left}) and $|\fR|=10^{-6}$ (\textit{right}) cases. The bottom pair of panels shows the same radial acceleration relations as in the middle pair, but $\log_{10}(g_{\mathrm{tot}})-0.75\log_{10}(g_{\mathrm{bary}})$ is plotted instead of $\log_{10}(g_{\mathrm{tot}})$, so as to flatten the curves and spread them out for easier distinguishability. As discussed in the text, galaxies Au21 and AuL1 are highlighted in red in all panels. The dotted red line in the right-hand rotation curve panel shows the standard gravity rotation curves for Au21 and AuL1, to enable more direct comparison with the $f(R)$ curves.}
    \label{F:All_RC_RAR}
\end{figure*}

The rotation curves displayed in the following section are calculated from post-processed simulation snapshots as follows.

Firstly, only particles in the disc plane are considered. In order to isolate these, the total angular momentum vector of the galaxy (within a sphere of radius 30 kpc around the centre) is calculated, and the disc plane is then the plane perpendicular to this vector. Particles within 0.5 kpc of this plane are admitted.

For each particle, the snapshot contains a standard gravity acceleration vector and a separate fifth force acceleration vector. Taking the inner product of the acceleration vector for a given particle (either including or excluding the modified gravity contribution) with the radial vector then gives an estimate of the square of the circular velocity at the particle position. The particles are then divided into radial bins, and the average circular velocity is calculated in each bin, giving the final rotation curve.

As a final note, the cosmological Compton wavelength in Hu \& Sawicki $f(R)$ gravity (for $n=1$) is approximately given by \citep{Cabre2012} 
\begin{equation}\label{E:Compton}
\lambda_C \approx 32 \sqrt{\frac{|\fR|}{10^{-4}}} \ \mathrm{Mpc}.
\end{equation}
On scales larger than this the fifth force would be suppressed even if the chameleon mechanism is not triggered. However, even for the smallest scalar field amplitude considered in this work, $|\fR| = 10^{-7}$, this wavelength is approximately 1 Mpc, well beyond the scales on which galaxy rotation curves can be measured. Outside the screening radius, fifth forces will hence affect the rotation curve out to the largest observed radii.

\section{Results}
\label{S:Results}

\subsection{Screening in Disc Galaxies}
\label{S:Results:Screening}

As described in \S\ref{S:Methodology:MG-GADGET}, we use the modified gravity solver from the \textsc{mg-gadget} code to post-process $z=0$ simulation snapshots from the \textsc{Auriga} project and calculate the scalar field $f_\mathrm{R}$ everywhere in the simulation volume. Figures \ref{F:FieldMaps} and \ref{F:Contours} show examples of the results of these calculations.

Figure \ref{F:FieldMaps} shows face-on and edge-on maps of the scalar field $f_\mathrm{R}$, calculated across planes going through the galaxy centres, for galaxies AuL8, Au13 and Au21, and for $|\fR|=8 \times 10^{-7}$. As in Figure \ref{F:Galaxies}, these three galaxies are chosen to represent a range of galaxy masses and morphologies. The resulting scalar field maps are qualitatively representative of the sample. In Figure \ref{F:FieldMaps}, it can be seen that in the outer regions of the galaxies (i.e. $R\gtrsim10$ kpc), the scalar field hovers roughly within an order of magnitude of the cosmic background value $\fR$. The scalar field in the innermost regions (i.e. $R\lesssim5$ kpc), however, is suppressed by many orders of magnitude, with values as low as $10^{-16}$. These regions are respectively the unscreened and screened regions.

In this figure, a sharp transition can be seen between these regions at $\sim5$ kpc for AuL8 and Au13, and $\sim15$ kpc for Au21, where $|f_\mathrm{R}|$ drops precipitously by many orders of magnitude. This is the screening radius of the galaxy, or more precisely its screening surface as there are clearly deviations from spherical symmetry. In the unscreened region outside the screening surface, particles are subject to a sizable fifth force, whereas in the screened region enclosed by the screening surface, the gradients of the scalar field are sufficiently small that the fifth force is suppressed, according to Equation (\ref{E:ScalarFieldForceLaw}). Equivalently, the ambient density in the screened region is sufficiently high, leading to an increased chameleon mass, which suppresses the range of the fifth force. 

Interestingly, the disc-shaped mass distribution of the galaxy is reflected in the shape of the $f_\mathrm{R}$ field, which appears to be compressed into the galactic disc plane as can be seen in the lower panels of Fig.~\ref{F:FieldMaps}. These effects can also be seen in the left-hand panels of Fig.~\ref{F:Contours}, where face-on and edge-on contour maps of the scalar field $f_\mathrm{R}$ for $|\fR|=10^{-6}$ are shown for Au20, overlaid on gas density projections. These findings reflect those of \citet{Burrage2015}, which analytically investigated the chameleon profiles around ellipsoidal objects.

As we shall see in the following sections, the location of the screening surface depends on a variety of factors: galaxy mass, galaxy density profile, environmental density, and $\fR$. 

The effect of changing $\fR$ can be seen in the right-hand panels of Figure \ref{F:Contours}. The screening surfaces of Au20 (or more precisely the intersection of the screening surface with a plane in or perpendicular to the disc plane) are shown as contours for three values of $|\fR|$: $5 \times 10^{-7}$, $8 \times 10^{-7}$, $1 \times 10^{-6}$, overlaid on gas density projections. Note that here, the screening surfaces are defined as the surfaces at which $f_\mathrm{R} = 10^{-4}\fR$. This was found to consistently fall inside the narrow transition zone. In this figure, it can be seen that larger values of $|\fR|$ correspond to smaller screening radii, and vice versa. For stronger background amplitudes of the scalar field, i.e. $|\fR| \gtrsim 2 \times 10^{-6}$, all galaxies investigated are entirely unscreened. Conversely, for most galaxies, weaker values ($\sim10^{-7}$) lead to screening radii beyond the range over which observed rotation curves are typically measured.

\subsection{The Effect of Chameleon f(R) Gravity on Galaxy Rotation Curves and Radial Acceleration Relations}
\label{S:Results:RCs}

The presence of a screening surface and the emergence of a fifth force on its outside have an impact on the dynamics of galaxies. Figures \ref{F:Au9_RC_RAR} and \ref{F:All_RC_RAR} display the resulting effects on the rotation curves and radial acceleration relations of the studied galaxies. We make the assumption that the disc is rotationally supported everywhere. 

The left-hand panel of Figure \ref{F:Au9_RC_RAR} shows the rotation curves calculated in the disc plane, including the fifth force contribution, of Au9 for six different values of $|\fR|$: $10^{-7}$, $5 \times 10^{-7}$, $8 \times 10^{-7}$, $10^{-6}$, $2 \times 10^{-6}$, and $10^{-5}$. In each case, the rotation curve beyond the screening radius\footnote{Note that in the remainder of this work, screening radius refers to the distance of the screening surface from the galaxy centre in the disc plane.} is enhanced by the additional presence of the fifth force. 

As was found in the previous sub-section, the cases $|\fR| = 2 \times 10^{-6}$ and $10^{-5}$ correspond to the galaxy being entirely unscreened and the rotation curve being enhanced with respect to the standard gravity rotation curve throughout the galaxy. Conversely, the case of $|\fR| = 10^{-7}$ gives a screening radius larger than the range shown in the plot, and outside the typical range spanned by observed rotation curves. Thus, the predicted rotation curve is identical to that of standard gravity.

The intermediate cases, however, are the most interesting. At the screening radius, the rotation curve shows a marked kink or `upturn', as it transitions between the screened and unscreened regimes.

Studying the galaxies of the SPARC sample \citep{Lelli2016}, \citet{McGaugh2016} found a remarkably tight relation between the total acceleration at each point inferred from rotation curves, and the acceleration due to baryonic mass inferred from observed light distributions. We have studied the effect of chameleon $f(R)$ gravity on this `radial acceleration relation'. The results of this can be seen in the right-hand panel of Figure \ref{F:Au9_RC_RAR}. In this figure, the baryonic acceleration $g_\mathrm{bary}$ is calculated at a given radius assuming spherical symmetry and neglecting a fifth force, i.e. adopting $\frac{GM_\mathrm{bary}(<R)}{R^2}$, while the total acceleration $g_\mathrm{tot}$, which would be measured from rotation curves, is calculated from the actual gravitational accelerations of the simulation particles, including the fifth force contribution.

Observed radial acceleration relations are typically smooth curves with only a small upward curvature. If no screening radius is present in the considered range, the simulation predictions give almost straight lines in this log-log plot. Mirroring the rotation curves in the left-hand panel, the strongest $\fR$ values give lines that are consistently enhanced with respect to the standard gravity case, while at the other end, the $|\fR|=10^{-7}$ line is identical to the standard gravity case. The intermediate cases, meanwhile, show marked bumps in the relations, corresponding to the screening radii. This is a promising result: the absence of these easily distinguishable bumps in observed radial acceleration relations could place strong constraints on $\fR$.   

Also shown in the right panel of this figure is the best-fit function for the radial acceleration relation from the SPARC galaxies \citep[equation 4 from][]{McGaugh2016}. As can be seen here with the case of Au9, the \textsc{auriga} galaxies are consistently to be found lying above this best-fit relation. This mirrors the findings of other simulations, e.g. the MassiveBlack-II simulations \citep{Tenneti2018}, and could be related to the central density profile of the dark matter halo.

Figure \ref{F:All_RC_RAR} shows the rotation curves and radial acceleration relations of all thirteen galaxies studied, for $\Lambda$CDM and $|\fR|=10^{-6}$. Galaxies Au21 and AuL1 are in red, in order to highlight two contrasting cases: the upper rotation curve, that of Au21, shows a clearly distinguishable upturn at its screening radius, $\sim 10$ kpc, in the upper right panel, while AuL1 is unscreened throughout, yielding no such observable feature.

It is also worth noting that in some cases, the upturns occur at locations where the rotation curve is already steeply rising. The curves are thus much smoother and the feature potentially less distinguishable observationally. However, the corresponding radial acceleration relations, seen in the central panels of Figure \ref{F:All_RC_RAR}, retain a marked bump.

The radial acceleration relations in the lower panel of the figure have been flattened for greater distinguishability.

\subsection{The Coupling of Stars to the Fifth Force}
\label{S:Results:Stars}

\begin{figure}
    \includegraphics[width=\columnwidth]{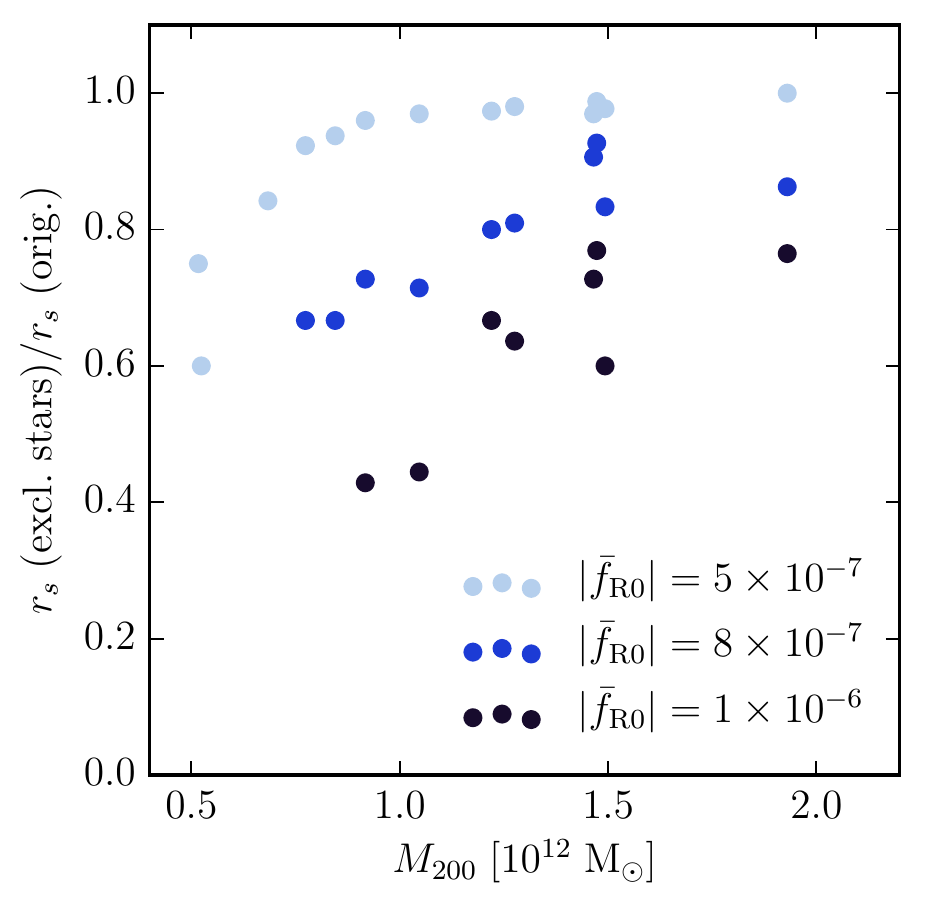}
    \caption{Ratio of screening radii including and excluding star particles as a source of the fifth force, as a function of halo mass. The three colours indicate calculations for three different values of $\fR$. For smaller $\fR$ values, the screening radius is larger and falls in a region in which the density is dominated by dark matter so that the effect of including / excluding stars is reduced. Therefore, the screening is less sensitive to self-screening of stars in this regime.}
    \label{F:Stars_Rs_M}
\end{figure}

\begin{figure*}
    \includegraphics[width=2\columnwidth]{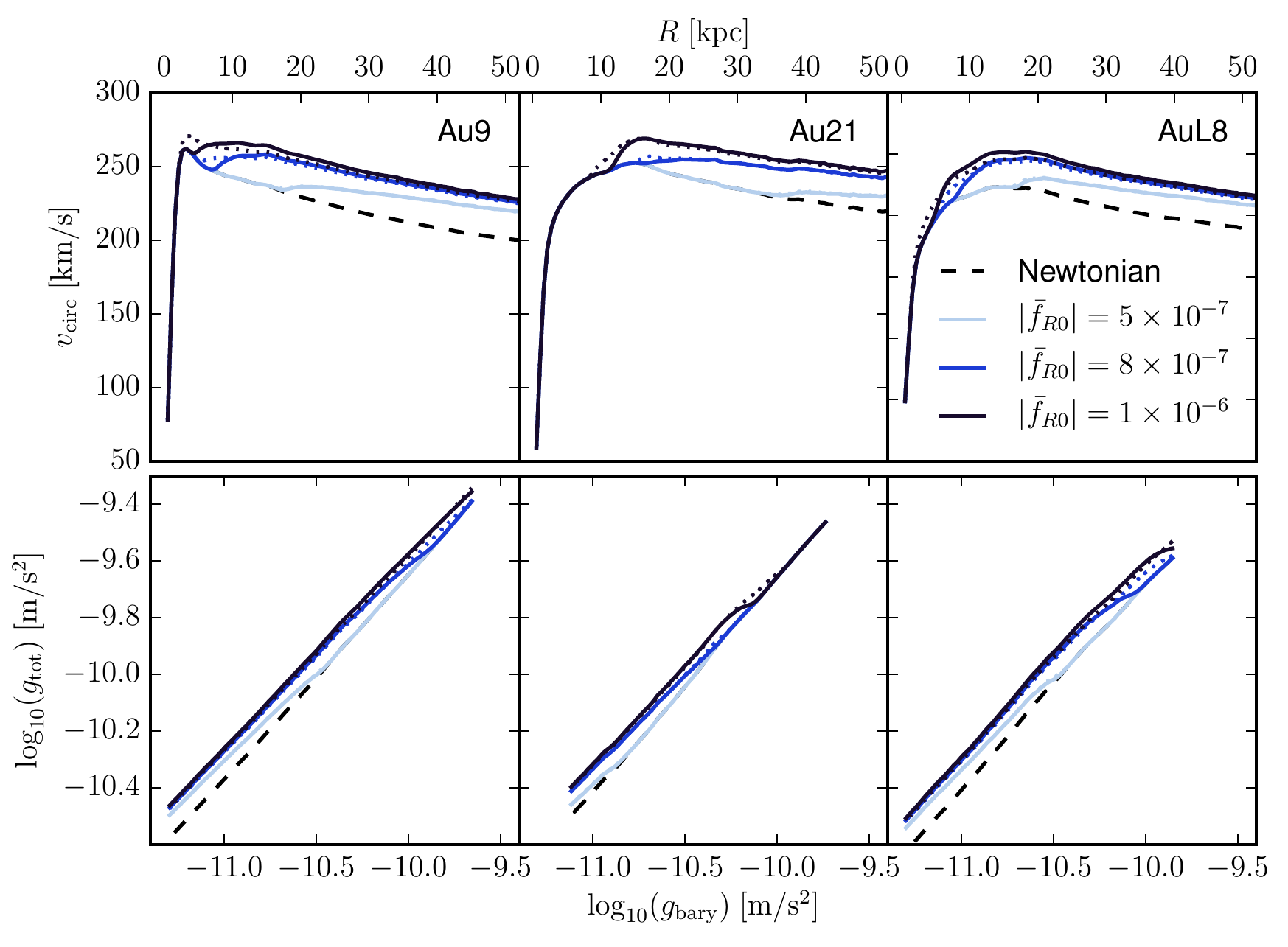}
    \caption{Rotation curves (\textit{top}) and radial acceleration relations (\textit{bottom}) for the gas components of Au9, Au21, and AuL8, including (\textit{solid}) and excluding (\textit{dotted}) the contribution of star particles to the fifth force. The dashed lines ignore the fifth force contributions altogether and hence correspond to $\Lambda$CDM.}
    \label{F:Stars_RC_RAR}
\end{figure*}

Like all cosmological simulations, modified gravity N-body simulations have limited spatial and mass resolution. Compact bodies such as stars are not resolved in cosmological runs, and are instead treated using simulation particles that represent whole stellar populations.

Ordinarily, this is a sufficient approximation, but in the case of screened modified gravity theories, one might expect some or all of the stars to be sufficiently dense so as to be self-screened, and therefore neither source nor couple to the fifth force, depending on the environment of a given star and the background amplitude of the scalar field \citep{Davis2012}.

One effect of this would be that for a given galaxy, the stellar rotation curve would differ from the gaseous rotation curve; the latter would show an upturn absent in the former. Alternatively, if a galaxy is completely in the unscreened regime, a difference in the normalisation of the two rotation curves would be observed (see, e.g., the $|\fR| = 10^{-5}$ model in the left panel of Fig.~\ref{F:Au9_RC_RAR}). \citet{Vikram2014} searched for evidence of such a different normalisation in observed rotation curves of isolated dwarf galaxies. No significant effect was observed and they were able to derive a constraint of $|\fR|<10^{-6}$ from this. 

Another effect would be that the fifth force in a given galaxy would be expected to be somewhat weaker than those calculated in the previous subsections if stars do not act as sources of the fifth force. In order to quantify this effect, we have trialled an alternative method for post-processing the \textsc{auriga} simulations. The star particles in the snapshots are included in the calculation of the standard gravitational acceleration, but excluded from the calculation of the scalar field. This corresponds to the extreme scenario, in which every star is fully self-screened, and thus not coupling to the scalar field. In reality, the scalar field would be an intermediate between this solution and the original solution of the previous subsections. In particular, for values of $|\fR| \geq 10^{-6}$ we would expect stars to start becoming unscreened and thus recovering our original solution \citep{Davis2012}. However, performing this fully self-consistent calculation would be difficult in practice and we choose instead to show the maximal effect of stellar self-screening.

Figures \ref{F:Stars_Rs_M} and \ref{F:Stars_RC_RAR} show the results of this test. Figure \ref{F:Stars_Rs_M} shows, for three different $\fR$ values, the ratio of the screening radii for all 13 galaxies in the new and original solutions, as a function of halo mass. One immediately apparent point is that the effect of excluding the stars from the scalar field calculation results in a shrinking of the screening radius. This is to be expected, as the exclusion of stars results in an object of lower \textit{effective} mass, that has a correspondingly smaller screening radius.

Another point is that higher amplitudes for the background scalar field result in greater differences between the two solutions. The reason for this is that, as seen in Figure \ref{F:Contours}, higher values for $|\fR|$ give smaller screening radii. At smaller radii, the stellar population becomes an increasingly dominant component of the overall density profile. Indeed, it is typically the case in the \textsc{auriga} galaxies that the stars dominate over the dark matter component in the central few kpc, and are a significant component for a few kpc beyond. Thus, the effect of excluding them from the scalar field calculation becomes more significant towards the centre. It is also for this reason that lower mass haloes appear to show more of a difference; the lower mass haloes tend to have smaller screening radii than their larger counterparts for a given $\fR$.

Figure \ref{F:Stars_RC_RAR} shows the effect of stellar self-screening on the rotation curves and radial acceleration relations of the gas component for three galaxies, and three values of $\fR$. As in Figure \ref{F:Stars_Rs_M}, the effect is that of shrinking the screening radii, and thus shifting the locations of the rotation curve upturns and the corresponding bumps in the radial acceleration relations. As before, the magnitude of the shift varies with the galaxy in question, and the value assumed for $\fR$. In the most extreme case, the shift in the rotation curve upturn is 2-3kpc. This is a sizable shift, but it should be borne in mind that this is the most extreme shift caused by the most extreme assumption: that all stars are fully self-screened, and that the screening radius of the galaxy is at a position where the density has a large stellar contribution. To test gravity, it might be preferable to use rotation curves at large radii, where the stellar density contribution is small. Then the effect of stellar self-screening on the rotation curve of the gas component is small as well, as seen for the $|\fR| = 5 \times 10^{-7}$ model.

\subsection{The Effects of Environmental Screening}
\label{S:Results:Environment}

\begin{figure*}
    \includegraphics[width=2\columnwidth]{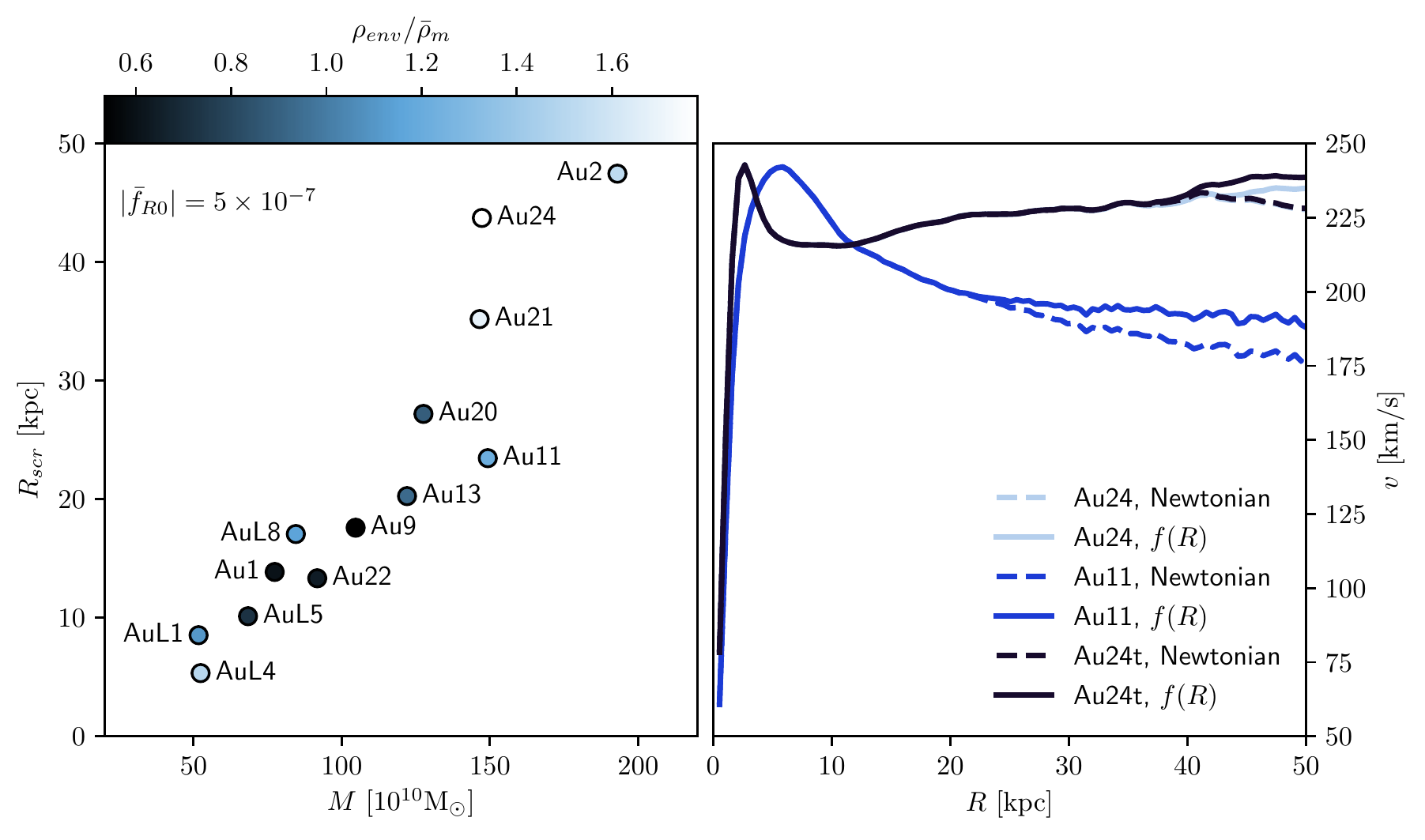}
    \caption{\textit{Left:} Screening radius as a function of mass for all 13 galaxies, calculated with $|\fR|=5 \times 10^{-7}$. The colours of the circles indicate the `environmental density' of the galaxy: the average density in a spherical shell centred on the galaxy, with inner radius $5R_{200}$, and outer radius $20R_{200}$, in units of the cosmic mean matter density. \textit{Right:} For the same value of $\fR$, rotation curves for Au11 and Au24. The `Au24t' result was obtained by transplanting Au24 into the environment of Au11 as discussed in the text. Solid lines include the fifth force contribution, while the dashed lines ignore it. Au24 and Au11 were chosen for this test because of their similar masses but differing environmental densities (see left panel).}
    \label{F:Environment}
\end{figure*}

The environment of a galaxy is expected to play a role in its screening. For example, a galaxy situated in a dense, group environment should have a larger screening radius than that of a galaxy situated in an underdense void. 

One can conceptualise this in the following way: from Equations~(\ref{E:FieldEOM}) and (\ref{E:ModifiedPoisson}) one finds that for small perturbations in the unscreened (low-curvature) regime the scalar field perturbation is given by $\delta f_R \equiv f_R - \bar{f_R} \approx -2/3 \, \phi_{\rm N}$, where $\phi_{\rm N}$ is the Newtonian gravitational potential of the considered mass distribution. Screening is triggered approximately when $\delta f_R \approx -\bar{f_R}$, as then $f_R \approx 0$ and its gradient which controls the fifth force is also suppressed. Thus, the screening becomes effective roughly when $|\phi_{\rm N}| \gtrsim 3/2 \, |\bar{f_R}|$ \citep[see, e.g.,][]{Hui2009, Arnold2014}. If an overdense environment contributes to the Newtonian potential at the position of a halo, the halo's own potential does not then have to be as deep to trigger screening.

Figure \ref{F:Environment} shows the results of a preliminary investigation into the magnitude of this effect. The left-hand panel shows the relation between screening radius and halo mass for all 13 galaxies in our sample, for $|\fR|=5 \times 10^{-7}$. The colours of the filled circles encode the 'environmental density': the average density in a spherical shell centred around the galaxy, with inner radius $5R_{200}$, and outer radius $20R_{200}$. The inner radius was chosen as we found that for smaller radii the density profiles typically still follows the NFW profile, while the outer radius is comparable to the cosmological Compton wavelength, so that the effect of the environment on even larger scales on $f_R$ should be suppressed. 

The left panel of Figure \ref{F:Environment} indeed suggests that there is a mild tendency of halos in dense environments having larger screening radii at similar halo mass. 

Taking two galaxies with similar masses but very different environmental densities, Au24 and Au11, we investigated the effect of transplanting the former into the environment of the latter. A sphere of radius $5R_{200}$ was cut out from the simulation volume of Au24 and placed into that of Au11, in the place formerly occupied by the galaxy Au11. This modified snapshot was then post-processed with \textsc{mg-gadget} as usual.

The results of this test are shown in the right-hand panel of Figure \ref{F:Environment}, which shows the rotation curves of the three galaxies: the original Au11 and Au24, as well as the transplanted Au24, referred to as `Au24t'. As in the left-hand panel, $|\fR|=5 \times 10^{-7}$ was assumed. The effect of the differing environments on the scalar field profiles is marginal. Only a small part of the difference in screening radius between Au11 and Au24 is explained by the environmental density. At least in this case, differences in the self-screening, due to different halo concentrations, largely dominate over differences in the environmental screening. 

A major factor in the limited effect of environmental screening is likely that the galaxies of the \textsc{auriga} sample are isolated galaxies; this was a criterion for selecting the galaxies from the parent simulation volume. Environmental screening would likely play a much larger role in the scalar field profiles of cluster galaxies.

When performing observational tests of screened modified gravity with galaxy kinematics, isolated galaxies would also be preferable to those inhabiting dense environments, so as to reduce the degeneracy between environmental screening and background field amplitude. The \textsc{auriga} simulation sample should hence be a suitable testbed for such studies. Residual environmental effects could likely be further mitigated by taking an estimate of the density of a galaxy's environment into account when using its rotation curve to constrain modified gravity.

\section{Discussion}
\label{S:Discussion}

\begin{figure}
    \includegraphics[width=\columnwidth]{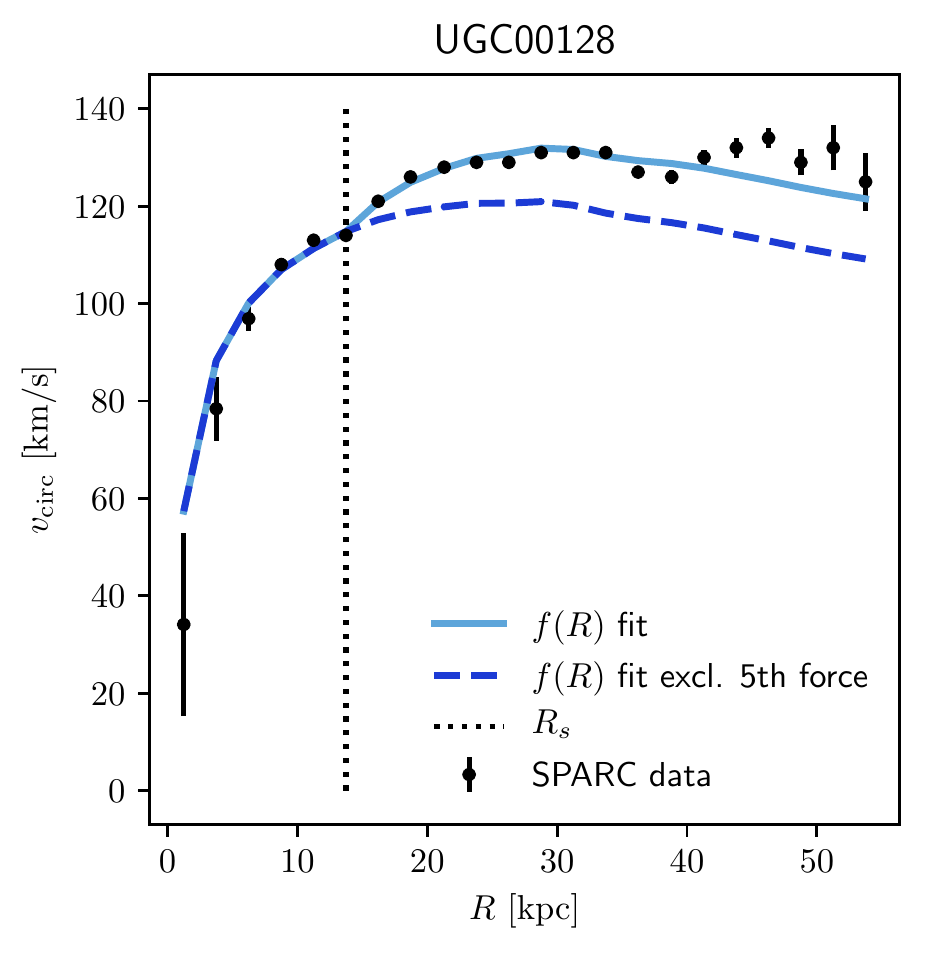}
    \caption[SPARC]{Rotation curve of galaxy UGC00128 created using data from the SPARC sample \citep{Lelli2016}, along with a fit created by a similar method as in \citet{Katz2017}, but additionally incorporating $f(R)$ gravity by fitting the screening radius $R_s$ (indicated by the \textit{vertical dotted line}).}
    \label{F:SPARC}
\end{figure}

We have found in section \S\ref{S:Results:RCs} that in $f(R)$ gravity galaxy rotation curves exhibit a distinct feature near the screening radius, namely an upturn in the circular velocity. A corresponding bump can be seen in the radial acceleration relation. Here, we discuss the utility of these features to constrain modified gravity models. 

While features of the predicted size would be clearly visible in measured rotation curves, great care must be taken in their interpretation. In particular, a potential astrophysical origin of such features as, e.g., a substructure in the galaxy or a complex dynamical state after a merger have to be ruled out.

However, while such astrophysical effects might cause such features in individual rotation curves, chameleon screening would consistently produce them in all galaxies in which the screening radius falls in the radial range where the rotation curve is measured. Furthermore, for a given density profile, the radius at which the feature should occur can also be predicted. This should in principle allow a robust interpretation. Before further elaborating on this, it is worth having a brief look at some observational rotation curve data.

Some observed rotation curves do display upturns that are qualitatively similar to those predicted in this work \citep[see, e.g., the SPARC sample,][]{Lelli2016}. Figure \ref{F:SPARC} illustrates this. It shows the rotation curve of galaxy UGC00128 using data from the SPARC sample. UGC00128 is an isolated field galaxy, we do hence not expect a sizeable environmental screening effect. In this example it is indeed possible to fit the feature at $\sim 13$ kpc with an upturn induced by the transition from the chameleon-screened to the unscreened regime. This is shown by the light blue curve. This fit to the rotation curve was performed following the procedure employed in \citet{Katz2017}, modelling UGC00128 as a dark matter halo with a gaseous and stellar disc. The gaseous and stellar components are constrained observationally, assuming a mass-to-light ratio, while the halo is fit with an NFW profile \citep{Navarro1996}. In \citet{Katz2017}, three parameters are fit: the halo concentration parameter $c_\mathrm{vir}$, halo virial velocity $V_\mathrm{vir}$, and mass-to-light ratio $M/L$. In addition to this, we incorporate the effect of $f(R)$ gravity by introducing one additional parameter: the screening radius $R_s$. Beyond this radius, a simple spherically symmetric fifth force profile is assumed \citep[see, e.g. eq. (10) in][]{Sakstein2013}. The aforementioned upturn feature at 13 kpc is well reproduced by the effects of $f(R)$ gravity.

As discussed above, the presence of an upturn in a given observed rotation curve is not necessarily evidence for the presence of a fifth force, but could instead be a result of astrophysical effects. However, if it is indeed due to Chameleon $f(R)$ gravity, the position of the upturn implies a specific value for $\fR$, which in turn dictates the locations of possible features in the rotation curves of other galaxies. If features are consistently found at these locations, this would lend strong support to $f(R)$ gravity. Conversely, if features are absent at these radii, that would suggest the original upturn was due to some other effect. Thus, the strongest conclusions can be drawn by performing fits to large ensembles of rotation curves and testing whether a single $\fR$ value allows good fits for the whole sample. In this context, it is however worth considering the two potential complications that we have investigated in Secs.~\ref{S:Results:Stars} and \ref{S:Results:Environment}, i.e. stellar self-screening and environmental screening which could both affect the exact position of an upturn.  

The effect of stellar self-screening on the overall scalar field solution has been neglected in the bulk of the calculations throughout this work. This assumption is dropped in \S\ref{S:Results:Stars}, where we instead consider the opposite extreme, in which all stars are assumed to be fully self-screened, and not acting at all as source of the fifth force. This typically led to an inward shift of the screening radius, in the most extreme cases by 2-3kpc. A safer option to avoid uncertainties related to this might be to use observed rotation curves at large radii, where the stellar contribution to the mass density is subdominant and a potential shift of an upturn is much smaller.

In addition to potentially preventing stars from acting as a source of the fifth force, self-shielding would also (partly) prevent them from feeling the acceleration by the fifth force. This would result in different rotation curves for stars and gas, as the latter would not self-screen. Rotation curves of the gas component at larger radii may, hence, be the most reliable probe of a fifth force. Alternatively, one can search for differences between stellar and gas rotation curves. This approach has been adopted by \citet{Vikram2014}, who analysed the gaseous and stellar rotation curves of dwarf galaxies. Rather than searching for upturns, that work searches for differences in the normalisation between gaseous and stellar rotation curves, caused by stellar self-screening, in galaxies that are likely unscreened otherwise. 

In \S\ref{S:Results:Environment}, it was found that the environment of a galaxy also plays a role in determining the scalar field solution. Galaxies in overdense environments have screening radii further from their centre than they would have if the environmental density was equal to the cosmic mean. Conversely, void galaxies have screening radii closer in. These environmental effects will therefore also alter the position of upturns in rotation curves. In the case considered in \S\ref{S:Results:Environment}, the effect of environmental screening was nevertheless found to be marginal. However, Au24 and Au11 are among the most massive galaxies in the sample, and environmental screening might well play more of a role in lower-mass galaxies. The effect of environmental screening should therefore be accounted for when performing a quantitative fit to observed rotation curves. This could either be done statistically, with an additional free parameter for the environmental density with a prior informed by simulations, or by directly studying the environment of galaxies with measured rotation curves. A step in the latter direction is the `screening map' presented by \citep{Cabre2012}; a three-dimensional map, covering a large portion of the sky, which employs large galaxy and cluster catalogues in order to calculate, at each point, the Newtonian potential due to external objects. If, for a given astrophysical object, $\Phi_{ext} \ll |\fR|$, then one can assume that self-screening of the galaxy dominates over environmental screening. \citet{Vikram2014} have employed this screening map in their sample selection. Furthermore, \citet{Desmond2018} have built upon the above work by creating an updated screening map, featuring more sophisticated techniques and a more complete sky map.

As an alternative to searching for upturns, one could also use completely smooth sections of rotation curves to rule out the presence of a screening radius in those regions. Using such sections in a sample of galaxies of different masses should allow the ruling out of large parts of the modified gravity parameter space.

For our roughly Milky Way-sized halos we find that the galaxies are typically screened at the galactocentric distance of the Solar System of $\sim 8 \mathrm{kpc}$ for background field amplitudes of $|\fR| \lesssim 8 \times 10^{-7}$, suggesting that $|\fR|$ cannot be much larger than this value. Searching for evidence of screening in rotation curves at larger radii and/or lower mass galaxies will allow probing even smaller field amplitudes. In the SPARC sample, accurate rotation curves have been measured for objects with rotational velocities down to $\sim 50$ km/s, roughly four times smaller than that of the Milky Way. The square of the circular velocity is expected to be roughly proportional to the depth of the Newtonian potential. The latter can be used to estimate the maximum background field amplitude at which Chameleon screening is triggered in an object (see Sec.~\ref{S:Results:Environment}). Together, this suggests that using rotation curves of lower mass galaxies, it should be possible to constrain $f(R)$-gravity down to $|\fR|$ values of $\sim  10^{-7}$, which would be very competitive compared to other techniques.

\section{Summary and conclusions}
\label{S:Conclusions}

We have studied the impact of Chameleon-$f(R)$ gravity on rotation curves and radial acceleration relations of disc galaxies. To this end, we have post-processed state-of-the-art $\Lambda$CDM simulations of disc galaxy formation from the \textsc{auriga project} with the modified gravity solver of the \textsc{mg-gadget} code. This is numerically much cheaper than performing full physics, galaxy formation simulations with $f(R)$ gravity, which remains very challenging. The validity of this post-processing approach is established in Sec.~\ref{S:Methodology:MG-GADGET}.

In addition to investigating the kinematic structure, we have studied the scalar field morphology and the transition from the screened to the unscreened region. Our main findings are:

\begin{itemize}
\item In $f(R)$ gravity, the scalar field iso-contours in disc galaxies inherit an oblate shape from the mass distribution. This results in a discoid screening surface (rather than a simple screening radius). This needs to be taken into account when predicting modified gravity effects on rotation curves.
\item At the position where the galactic disc penetrates the screening surface, a distinct upturn is present in the rotation curve. The rotational velocity in the unscreened region is enhanced by the fifth force by up to a factor $\sim\sqrt{4/3}$.
\item A corresponding distinct bump is present in the radial acceleration relation.
\item Lower values of the comic background scalar field, $|\fR|$, lead to larger screening radii, and therefore rotation curve upturns that are more distant from the galactic centre. Conversely, more massive objects have smaller screening radii at fixed $|\fR|$.
\item Stellar self-screening and environmental screening can also affect the position of the upturn in the rotation curve. The former effect is negligible for upturns at large radii where the stellar contribution to the mass density is small. Environmental screening is a sub-dominant effect for the Milky Way-sized galaxies considered here, may however be more important in lower mass galaxies.
\item Stellar self-screening will also result in different rotation curves of stars and gas. Since the gas will not self-screen, its rotation curve might be easier to interpret.
\item The predicted rotation curve upturns are qualitatively similar to upturns seen in at least some observed galaxies. These signatures, provide a potentially promising avenue toward strong constraints on $f(R)$ gravity. However, as discussed in Sec.~\ref{S:Discussion}, a careful statistical analysis of galaxy samples will be needed to unambiguously distinguish modified gravity effects from astrophysical effects on rotation curves and radial acceleration relations.
\item In the model with the smallest background scalar field amplitude, $|\fR| = 1\times 10^{-7}$, the rotation curves and radial accelerations of all galaxies considered were indistinguishable from the $\Lambda$CDM case. Conversely, at the other end of the spectrum, all galaxies were unscreened for $|\fR| = 10^{-5}$ and $2 \times 10^{-6}$, and therefore their rotation curves were enhanced with respect to $\Lambda$CDM throughout the entire disc. The intermediate values of $\fR$ are hence the most interesting. It is at these values that rotation curves and radial acceleration relations display upturns and bumps that would be visible in observational data. Note, however, that lower mass galaxies will be sensitive to correspondingly smaller $|\fR|$ values. Sensitivities down to $|\fR| \sim  10^{-7}$ should be achievable with existing data.
\end{itemize}

Our results indicate that rotation curves and radial acceleration relations can provide constraints on screened modified gravity that are very competitive with constraints from larger scales, e.g., employing the next generation of galaxy clustering and weak lensing surveys. Furthermore, they are complementary to these surveys as they test gravity on different scales. Applying this technique to data seems hence a very promising direction to pursue and we plan to do so in future work.

\section*{Acknowledgements}

We would like to thank Volker Springel, Debora Sijacki, Claudio Llinares, Harley Katz and Vasily Belokurov for helpful comments and discussions. We would also like to thank the \textsc{auriga project} team for allowing us access to their simulations.

APN thanks the Science and Technology Facilities Council (STFC) for their PhD studentship. EP acknowledges support by the Kavli Foundation. ACD is partially supported by STFC Consolidated Grants No. ST/P000673/1 and No. ST/P000681/1. CA acknowledges support from the European Research Council through grant ERC-StG-716532-PUNCA.

This work used the DiRAC (www.dirac.ac.uk) system: Data Analytic at the University of Cambridge [funded by BIS National E-infrastructure capital grant (ST/K001590/1), STFC capital grants ST/H008861/1 and ST/H00887X/1, and STFC DiRAC Operations grant ST/K00333X/1]. DiRAC is part of the National E-Infrastructure.


\bibliographystyle{mnras}
\bibliography{library}

\bsp	
\label{lastpage}
\end{document}